# Charge Transport in Dendrimer Melt using Multiscale Modeling Simulation

Saientan Bag[1] , Manish Jain[1] and Prabal K Maiti[1*]

[1]Center for Condensed Matter Theory, Department of Physics, Indian Institute of Science, Bangalore 560012, India

*Abstract*— **In this paper we present a theoretical calculation of the charge carrier mobility in two different dendrimeric melt system (Dendritic phenyl azomethine with Triphenyl amine core and Dendritic Carbazole with Cyclic Phenylazomethine as core), which have recently been reported[1] to increase the efficiency of Dye-Sensitized solar cells (DSSCs) by interface modification. Our mobility calculation, which is a combination of molecular dynamics simulation, first principles calculation and kinetic Monte Carlo simulation, leads to mobilities that are in quantitative agreement with available experimental data. We also show how the mobility depends on the dendrimer generation. Furthermore, we examine the variation of mobility with external electric field and external reorganization energy. Physical mechanisms behind observed electric field and generation dependencies of mobility are also explored.**

*Email: maiti@physics.iisc.ernet.in, Phone: (091)80-2293-2865.

## Introduction

Organic semiconductors are the building blocks of the various optoelectronics devices such as solar cells, light emitting diodes (LED), field effect transistor (FET) etc[2]. Charge transport properties of the constituent semiconductor play an important role[3] in the efficiency of the optoelectronic devices. Exploration of new organic semiconducting material with high and tunable charge carrier mobility (for high performance of optoelectronic device), is an active area of research for past several years . The charge carrier mobility of the disordered organic material reported so far usually lies in the range[4-8] of $10^{-7}$ to $10^{-10}$ cm$^2$V$^{-1}$S$^{-1}$ . On the other hand the organic semiconducting crystals can give rise to mobility as high[9-10] as 0.1 cm$^2$V$^{-1}$S$^{-1}$ . Although mobility is very high in comparison to the disordered semiconductors the morphology of the organic crystals limits their applicability to a large class of optoelectronic devices. For application to a specific device one needs to have organic semiconductor[11] with specific morphologies. For instance, in some applications like OLEDs, solar cell etc. conduction both parallel and perpendicular to the substrate[11] is required. Organic crystals fail to provide isotropic mobility which is required in this case. In addition, semiconducting crystals are very unrealistic for use in large scale application owing to the difficulty and cost of producing single crystals on a large scale. Cheaper fabrication techniques, such as vacuum sublimation and solution processing, always results in disordered film[2]. Despite the low mobility, the disorder organic semiconducting material is still a promising candidate for the application in large class of optoelectronical devices. The amorphous dendrimer melt system, a disordered organic semiconductor, is a especially good choice for these applications because of its well defined molecular weight[12-13] , batch-to-batch reproducibility and high purity synthesis. Understanding the charge carrier mobility in disordered semiconductors like dendrimers is a step forward in the realization of organic electronics.

The charge carrier mobility in the nitrogen center stilbene dendrimer melt system reported earlier[11, 13-15] lies between $10^{-3}$ to $10^{-9}$ cm$^2$V$^{-1}$S$^{-1}$ depending on the structure and chemical composition of the respective dendrimer. In most of the theoretical work Gaussian disordered model[11, 14, 16-17] was used with empirical fitting parameters. Gaussian disorder model is a macroscopic description of the charge transport phenomena where effect of microscopic details of morphology on mobility is not taken into account. As previously mentioned, the charge carrier mobility in disorder organic semiconductors, like dendrimer, strongly depends on the molecular structure, chemical composition and the packing of the molecules[12, 18-19] in a melt system. In a recent report M. E. Kose et. al[12] have presented a theoretical calculation of mobility in dendrimer using Monte Carlo simulations. In their study, Kose et al. used phenyl-cored four-arm thiophene dendrimers for the charge transport study and reported a charge carrier mobility in the range of $10^{-3}$ cm$^2$V$^{-1}$S$^{-1}$ which was two order of magnitude higher than the experimentally measured mobility.

Both theoretical[12] and experimental[15] study of generation and electric field dependence of the mobility in dendrimer has been reported in past. The reduction of mobility with the increase in the electric field strength has been seen in experiment and also in theoretical calculations. In contrast, there are conflicting views as to how the mobility depends on the dendrimer generation. In some oligomeric based conjugated dendrimers, mobility decreases almost one order of magnitude as one goes from G2 to G3[13], while an increase in mobility is seen in phenyl-cored thiophene dendrimer[12]. However, the physical mechanism behind the observed electric field and generation dependence of the charge carrier mobility is still not well understood.

In a recent report Nakashima et. al[1] have shown how the efficiency of the Dye-Sensitized solar cell (DSSCs) can be increased using the Carbazole dendrimer (with Cyclic Phenylazomethine as core and Dendritic phenyl azomethine with Triphenyl amine core) as the coating material to the TiO2 interface. Realizing the potential of these dendrimers, it is necessary to have the better understanding of the charge transport properties of these dendrimers.

In this paper we calculate the charge carrier (hole and electron) mobility of the two different dendrimer systems: Dendritic phenyl azomethine with Triphenyl amine core (Dpa-Tpa) and Dendritic Carbazole with Cyclic Phenylazomethine as core (Cpa-Cz). We calculate the charge carrier mobility in both these systems starting from a very realistic morphology which is obtained using fully atomistic molecular dynamics simulation. Our calculation produces the value of mobility which is of the same order of magnitude as measured in experiment[13] .We also study the electric field and the generation dependence of the mobility for both the dendrimer cases. Effect of external reorganization energy on the mobility is also reported. We propose the possible physical mechanism responsible for this kind of dependence and demonstrate this by theoretical calculation.

The paper is organized as follows: In the first section we discuss the details of the molecular dynamics (MD) simulation carried out to obtain a realistic morphology of the system through which charge transport will be simulated. In section II, we discuss the charge transport simulation using semi classical Marcus-Hush formalism. Finally, in section III we summarize the results and end with concluding remarks.

I. **MOLECULAR DYNAMICS SIMULATION**

## I.A.   MODEL AND COMPUTATIONAL DETAILS

The molecular structure of Dendritic Carbazole with Cyclic Phenylazomethine (Cpa-Cz) as core is shown in figure 1 for two different generations: generation 2 (G2) and generation 3 (G3). Figure 2 shows the molecular model of the  phenyl azomethine dendrimer with Triphenyl amine as core (Tpa-Dpa).  With increase in dendrimer generation, branching increases exponentially. Here we briefly outline the procedure followed to build the initial 3-d model of the dendrimers of both the above mentioned types. First we geometry optimize the core and the branch in Gaussian09 with HF/6-31G(d) as the basis set. During optimization, charges were calculated using electrostatic potential (ESP) method. Then the optimized fragments were added together using Dendrimer builder ToolKit[20] to get the 3-d model of the full molecule for further studies. The Antechamber[21] module of the AMBER[21] package with the GAFF[22] force field and ESP charge (calculation of charge fitted to electrostatic potential) was further used to prepare the system for simulation.

Four different systems were simulated: Dpa-Tpa for generation 2 (G2) and generation 3 (G3) and Cpa-Cz for generation 2 and 3. In all four cases, a system of 64 molecules was simulated. Initially the molecules were arranged in 16 columns of 4 molecules in each column. The columns were arranged in a square lattice. The lattice constant was chosen to get maximum possible density. The snapshots of the initial systems are shown in figure 3(a) and 3(b) .

After an initial energy minimization, the system was heated slowly from 0 K to 300 K at a constant pressure of 1 bar. The system was then subjected to simulated annealing of five repeat cycle from 400K to 600K and back, in steps of 50K, allowing 2ps in each step under NPT condition to release the molecules from potential traps or initial biases[23]. Finally, a  production run of 100 ns was performed at 300 K and at a pressure of 1 bar. At 300 K the system is in the amorphous melt state.  Temperature and pressure were kept fixed using  Berendsen weak temperature coupling and pressure coupling method[24] using a temperature coupling constant of 1 ps and pressure coupling constant of 2 ps.  During the NPT run, simulation box angles were fixed at $90^0$  and length of the simulation box was allowed to very independently in each direction. The bonds involving hydrogen were kept constrained using the SHAKE algorithm.  This allowed us to use an integration time step of 2 fs. All calculations were performed with parallel version of PMEMD[21].

## I.B.  RESULTS AND DISCUSSION

A representative equilibrated simulation snapshot of the system in case of Cpa-cz,G2 with 64 molecules T=300K is shown in figure 3(c) and 3(d). It shows a completely isotropic assembly of the dendrimers with a density close to ~1 gm/cc.

To extract the information about the structure the dendrimers after the equilibration we calculate the radius of gyration ($R_g$) of the dendrimer in all four cases. The numerical values of the radius of gyration (averaged over last 10 ns MD run) are given in Table 1. We see that the $R_g$ increases with increase in dendrimer generation. This nature is very similar to other classes of dendrimer like PAMAM[25], PETIM[26] and PPI[27].

TABLE 1: Average radius of gyration ($R_g$) of the Dpa-Tpa, and Cpa-Cz dendrimers for different generations.

| Dendrimer | Generation | $R_g$ (Å) |
|---|---|---|
| Cpa-cz | G2 | $11.80 \pm 0.06$ |
| Cpa-cz | G3 | $13.84 \pm 0.08$ |
| Dpa-Tpa | G2 | $15 \pm 0.07$ |
| Dpa-Tpa | G3 | $16.32 \pm 0.10$ |

## II. CHARGE CARRIER MOBILITY

### II. A. METHODOLOGY

To study the charge carrier mobility through the amorphous melt phase we follow the semi-classical Marcus-Hush formalism[28] which has been successfully used previously to predict the charge carrier mobility in wide variety of systems including organic semiconductors. Charge transport is described as thermally activated hopping mechanism of charge carriers between charge hopping sites. According to the formalism, the charge transfer rate $\omega_{ij}$ from $i^{\text{th}}$ charge hopping site to the $j^{\text{th}}$ hopping site is given by

$$\omega_{ij} = \frac{|J_{ij}|^2}{\hbar} \sqrt{\frac{\pi}{\lambda kT}} \exp\left[-\frac{(\Delta G_{ij} - \lambda)^2}{4\lambda kT}\right]. \qquad (1)$$

$J_{ij}$ is the transfer integral, defined as

$$J_{ij} = \langle \phi^i | \hat{H} | \phi^j \rangle. \qquad (2)$$

Here $\phi^i$ and $\phi^j$ is the diabatic wave function localized on $i^{th}$ and $j^{th}$ site respectively. $\hat{H}$ is the Hamiltonian of the two sites system between which the charge transfer takes place. $\Delta G_{ij}$ is the free energy difference between two sites. $\lambda$ is the reorganization energy , $\hbar$ is the Plank's constant, k is the Boltzmann's constant, and T is the temperature.

The workflow[29-31]of our charge transport simulation is as follows: i) We use fully atomistic MD simulation as described above to get the equilibrium morphology of the system ii) With this morphology we partition the system in charge hopping sites and decide the charge hopping pairs (pairs between which charge hopping is possible) iii) We calculate the transfer integral and all other quantities appearing in the rate expression for all charge hopping pairs. iii) Once the rate $\omega_{ij}$ is known for all charge hopping pairs, kinetic Monte Carlo[29-30, 32] method is used to simulate charge carrier dynamics and calculate the carrier mobility.

The details of molecular dynamics simulation has already been discussed in section I. With the equilibrium morphology in hand we partition the system in charge hopping sites. There is no strict rule in partitioning the system in charge hopping sites. Physical intuition[30] based on the localization of the wave function of a charge are often used to predict the hopping sites. To decide the charge hopping sites we look for the planar aromatic structure with rich $\pi$ electron cloud which is an ideal geometry for charge localization. In our case, the branches of the dendrimer form these planar structures. Two different branches always stay in different plane so they do not form the charge hopping sites together. So, we consider the cores and branches of the dendrimer as the charge hopping sites as shown in the figure 4. With this choice the number of charge hopping sites for single 2nd generation Cpa-Cz dendrimer turns out to be 10, while 3rd generation Cpa-Cz generates 22 hopping sites. After the partitioning, we get total 640 and 1408 hopping sites for G2 (Cpa-Cz) and G3 (Cpa-Cz) dendrimer system respectively. For Dpa-Tpa cases the number of hoping sites are same as CPA-Cz cases. To calculate the charge hopping pairs (pairs between which charge hopping is possible) we calculate first the distance between all atoms belonging to a hopping sites to all other atoms belonging to the other sites. In other words if there is *N* atom in a site and *M* atom in another site, total (*NM/2*) no of distances are being calculated. Now if any one of the calculated (*NM/2*) distances falls below a certain cut off, then the corresponding sites will be considered as hopping pairs. The reason behind using this kind of criteria to decide the charge hopping pairs is that the hoping sites are not points but occupy a finite portion of the space. A simple centre of mass to centre of mass distance calculation will be very unphysical in this case. The number of sites with which a particular site makes hopping pairs are defined as the neighbours to the particular site. In other words if a site has *n* number of neighbours then the charge can hop to *n* other sites from that particular site and the charge can come to that particular site from *n* other hopping sites. The highest occupied molecular orbital (HOMO) and lowest unoccupied molecular orbital (LUMO) are used[32-33] as diabatic wave function to calculate $J$ between the pairs for hole and electron transfer respectively. Figure 5 shows the HOMO and LUMO orbitals of the core and branch (hopping sites) of the two dendrimers. So $J$ is equal to $\langle HOMO^i|\hat{H}|HOMO^j\rangle$ and $\langle LUMO^i|\hat{H}|LUMO^j\rangle$ for hole and electron transfer respectively. We have not included HOMO-1 (LUMO+1), HOMO-2 ( LUMO+2 ) orbitals for calculating transfer integral for hole (electron) transport because the HOMO and LUMO orbitals of the charge hopping sites are not degenerate. To calculate these terms we use semi empirical Zerner's independent neglect of differential overlap (ZINDO) method[34] using Gaussian09 programme and the code available in VOTCA-CTP[30] module. The average values of the hole and electron transfer integrals for various cases are given in Table 2 below.

TABLE 2: Average values of the hole and electron transfer integrals for Cpa-CZ and DPa-Tpa dendrimers melt systems.

| Dendrimer | $J$ (Hole) | $J$ (Electron) | Dendrimer | $J$ (Hole) | $J$ (Electron) |
|-----------|-----------|----------------|-----------|-----------|----------------|
| Cpa-Cz (G2) | 0.13 eV | 0.018 eV | DPA-TPA (G2) | 0.44 eV | 0.01 eV |
| Cpa-CZ (G3) | 0.14 eV | 0.14 eV | DPA-TPA (G3) | 0.28 eV | 0.013 eV |

The reorganization energy $\lambda$ can be decomposed in two parts, inner sphere reorganization energy and outer sphere reorganization energy. Inner sphere reorganization energy takes care of the change in nuclear degree of freedom as charge transfer takes place from molecule $i$ to molecule $j$ and can be defined as

$$\lambda_{ij}^{int} = U_i^{nC} - U_i^{nN} + U_j^{cN} - U_j^{cC} \qquad (3)$$

Where $U_i^{nC}(U_i^{cN})$ is the internal energy of neutral (charged ) molecule in charged (neutral) state geometry and $U_i^{nN}(U_i^{cC})$ is the internal energy of neutral (charged) molecule in neutral (charged) state geometry. To calculate the terms appearing in equation 3, we first geometry optimized the hopping sites on neutral and charged state using Gaussian09 with B3LYP functional and 6-31g(d) basis sets. Then single point energy calculation with the optimized geometry and different charged state are performed to get the different terms in internal reorganization energy equation 3. Outer sphere reorganization energy is that part of the reorganization energy that takes into account the reorganization of the environment as the charge transfer takes place. The calculation of external reorganization energy is very much involved and intricate[30]. A rough estimate of the external reorganization energy is possible by considering the dendrimer melt system to be a dielectric medium. In that case the external reorganization energy can be expressed[35] by the following equation

$$\lambda_{ij}^{out} = \frac{e^2}{4\pi\varepsilon_0 R}\left(1 - \frac{1}{\varepsilon}\right) \qquad (4)$$

Here $e$ is the electronic charge, $\varepsilon_0$ and $\varepsilon$ are the dielectric constant of the vacuum and the dendrimer melt respectively. R is the effective radius of the charge hopping sites. For a value of R~15 Å and $\varepsilon$ in the range of 2-60 which is the typical value of $\varepsilon$ for the most polymeric systems[36-37], $\lambda_{ij}^{out}$ turns out to be in the range of 0.4 eV to 1 eV. In our calculation we do not calculate external reorganization energy exactly, but we calculate the mobility for different values of $\lambda_{ij}^{out}$ in the above range.

The free energy difference $\Delta G_{ij}$ appearing in the rate expression is the contribution[12] from the different sources as described below

$$\Delta G_{ij=}\Delta G_{ij}^{ext} + \Delta G_{ij}^{int} \qquad (5)$$

$\Delta G_{ij}^{ext}$ is contribution of external electric field defined defined $\Delta G_{ij}^{ext} = F.d_{ij}$. Here F is the external electric field and $d_{ij}$ is the displacement vector between the centre of mass of $i^{th}$ and $j^{th}$ hopping sites. $\Delta G_{ij}^{int}$ is the contribution in free energy difference due to different internal energies and can be written as

$$\Delta G_{ij}^{int} = U_i^{cC} - U_i^{nN} + U_j^{cC} - U_j^{nN} \qquad (6)$$

Where $U_i^{cC}(U_i^{nN})$ is the internal energy of site $i$ in charged (neutral) state and geometry. Now $\omega_{ij}$ is calculated for all hopping pairs and charge transport dynamics is simulated using kinetic Monte Carlo (MC) method . For kinetic MC, an in house code was developed with following algorithm[29-30]. We assign an unit positive charge (for hole transport) or negative charge (electron transport) to any of the hopping site $i$. At this point we initialize the time as $t = 0$. If there is N number of neighbour of the site $i$, then the waiting time $\tau$ of the charge is calculated according to the relation

$$\tau = -\omega_i^{-1}\ln(r_1) \qquad [\omega_i = \sum_{j=1}^{N}\omega_{ij}] \qquad (7)$$

and time is updated as $t = t + \tau$ . Where $j$ is the index of the neighbours (the hopping sites where the charge can hop from the site ) of the particular hopping site $i$ and $r_1$ is a random number between 0 to 1. To decide where the charge will hop among N neighbours we choose the biggest $j$ for which $\frac{\sum_j \omega_{ij}}{\omega_i} \leq r_2$. Here $j$ is the index of the neighbours of site $i$ and $r_2$ is a random number between 0 to 1. The above mentioned condition will ensure that the side $j$ is selected with probability $\frac{\omega_{ij}}{\omega_i}$. Now the position of the charge is updated and the above process is repeated.

The simulation was repeated with different initial position of charges (as in the snapshots obtained from MD). The charge carrier mobility along $x$ direction is determined[31] from the average charge velocity along that direction by

$$\langle v_x \rangle = \mu_x F_x \qquad (8)$$

$\langle v_x \rangle$ is the average charge velocity of the charge in $x$ direction, $\mu_x$ is the mobility along $x$ direction and $F_x$ is the applied electric field. Since the system is isotropic the mobility is same in all three directions. In our case average velocity was calculated from the $x$ component of unwrapped charge displacement (since periodic boundary condition was used) divided by total simulation time $t$.

### II.B. RESULTS AND DISCUSSION

We calculate the hole and electron mobility for both the dendrimer systems as a function of dendrimer generation (for both the G2 and G3 dendrimers). For the case of Cpa-Cz dendrimer, the hole mobility is of the order of $10^{-8}$ cm$^2$V$^{-1}$S$^{-1}$ while electron mobility is in the range of $10^{-12}$ cm$^2$V$^{-1}$S$^{-1}$ for an electric field of $10^5$V/cm for both G2 and G3.The magnitude of the electric field strength was chosen to match the experimental field strength[13] .On the other hand for the DPA-TPA dendrimer case, the electron mobility is of the order of $10^{-6}$ cm$^2$V$^{-1}$S$^{-1}$ and hole mobility is of the order of $10^{-8}$ cm$^2$V$^{-1}$S$^{-1}$ for the case of G2 dendrimer. For G3 dendrimer, both the hole and the electron mobilities are in the range of $10^{-13}$ cm$^2$V$^{-1}$S$^{-1}$. Our calculated mobility values are in quantitative agreement with the available experimental data[13]. Using time of flight measurement, Lupton et al.[13] reported hole mobility (majority carrier) values of ~$5 \times 10^{-8}$ cm$^2$ V$^{-1}$ S$^{-1}$ and ~$1.5 \times 10^{-8}$ cm$^2$ V$^{-1}$S$^{-1}$ in G2 and G3 conjugated dendrimer system respectively. The point to note from the above discussion is that for CPA-Cz case, hole mobility is higher than the electron mobility, while in DPA-Tpa case electron mobility is higher. To understand the observed difference in hole and electron mobility we calculate the HOMO and LUMO energy of core and branch of the dendrimers separately. The energy values are presented in Table 3.

TABLE 3: HOMO and LUMO energy of the core and branch of the Dpa-Tpa and Cpa–cz Dendrimer. Gaussian programme with B3LYP functional and 6-31G(d) basis set was used for energy calculation.

| Cpa-cz | HOMO ENRGY(In Hatree) | LUMO ENERGY(In Hartree) |
|--------|-----------------------|-------------------------|
| Core | -0.21125 | -0.05628 |
| Branch | -0.20002 | -0.02367 |

| DPa-Tpa | HOMO ENRGY(In Hatree) | LUMO ENERGY(In Hartree) |
|---------|-----------------------|-------------------------|
| Core | -0.17478 | -0.05415 |
| Branch | -0.23659 | -0.04545 |

For CPA-Cz dendrimer, HOMO energy of core and branch are very similar, while the LUMO energy are well separated. In the Marcus Hush formalism, the charge transport is described as hopping of hole (electron) through the HOMO (LUMO) of the charge hopping sites. A large difference in HOMO (LUMO) energies would necessarily indicate lower rate of hole (electron) transport between core and branches, since the rate depends exponentially on the energy difference. As a result, the overall mobility of electron will be low in comparison to hole (difference in HOMO energy of dendrimer core and branch is small) in melt system of CPA-cz . For the same reason the hole mobility is lower than the electron mobility in Dpa-Tpa dendrimer.

The point we want to emphasize here is that in case of very low mobility (approximately less than $10^{-10}$ $cm^2V^{-1}S^{-1}$) the hopping events are very rare and it is very difficult to get a steady state for the charge transport within reasonable simulation time. So the low mobility values may not be statistically significant. In the next section we will be discussing the electric field and external reorganization energy dependence of the mobility only for the cases where mobility value is of the order of $10^{-10}$ $cm^2V^{-1}S^{-1}$ or higher.

The variation of hole mobility of Cpa-cz dendrimer with respect to electric field is shown in figure 6. At low field the mobility (hole) is higher in G3 than in G2 , but as electric field increases the mobility in G3 decreases rapidly and after a certain value of electric field, mobility in G2 takes over the mobility in G3. The electric field dependence of hole and electron mobility for G2 Dpa-Tpa is shown in figure 7(a) and 7(b) respectively. The reason for the decrease in mobility with the increase in the electric field strength can be understood as follows: At low electric field, the charge chooses only the most favourable path in hopping process. As the field increases, the charge is forced to occupy also the sites which was previously very less favourable to hop on. As a result the movement of the charge is delayed[12] resulting decrease in mobility. The G3 dendrimer being more branched structure than G2 , allows more paths for the carrier to hop in comparison to G2. As a result under the application of electric field the mobility decreases at a faster rate in G3 than in G2. To have a better microscopic understanding of the electric field dependence of the mobility as described above we present a schematic diagram (Figure 8) representing few specific charge hopping sites and corresponding hopping probabilities for Cpa-cz (G3) dendrimer. At low electric field the charge always hops from site A to C (Figure 8) while at high electric field there is finite probability for the charge to hop to site C as well to site B.

At low electric field when the carrier is supposed to chose only the most favourable path, G3 has higher mobility (figure 6) than G2 in the case of CPa-cz dendrimer while in the case of DPA-Tpa dendrimer, carrier mobility is higher in G2 than in G3. To understand this generation dependence, we look into the possible hopping events transporting the charge from one place to another. Suppose $\omega_{core-core}$ is the average hopping rate between two cores and $\omega_{core-branch}$ is the average hopping rate between core and branch of the dendrimer. Now if $\omega_{core-core} \gg \omega_{core-branch}$, then charge transfer will mainly happen through the cores of the dendrimer as shown schematically in figure 9(a). As the dendrimer generation increases, the core-core distance between two dendrimer increases. To verify this we calculate the average distance between the centre of mass (COM) of the cores which are neighbours to each other. The COM distances between the core for all four cases are given in Table 4. We see that the average distance between the neighbouring cores increases with increase in generation.

TABLE 4: Average COM distance between the cores (core-core distance) which are neighbors to each other. The distances are calculated for both the G2 and G3 generations for both the Dpa-Tpa and Cpa-Cz dendrimers.

| Dendrimer | Generation | Core-Core distance (Å) |
|---|---|---|
| Cpa-cz | G2 | $10.16 \pm 1.5$ |
| Cpa-cz | G3 | $12.40 \pm 1.6$ |
| Dpa-Tpa | G2 | $12.48 \pm 1.8$ |
| Dpa-Tpa | G3 | $14.64 \pm 1.7$ |

As the core-core distance between two dendrimer increases, $\omega_{core-core}$ decreases, resulting overall decrease in mobility. On the other hand if $\omega_{core-core} \ll \omega_{core-branch}$, then charge will also come to branch and follow the kind of path shown schematically in figure 9(b). With the increase in generation, due to more branching, the more new paths will open up for the charge to chose. Among the new paths, some path will provide faster route and some will provide slower route for the carrier in comparison to the old paths. Since at low electric field the charge only chooses the most probable (fastest) path so as a whole the mobility will go up.

Finally, we study the variation of carrier mobility as a function of the external reorganization energy keeping the value in the range as discussed in section II.A. Figure 10 shows how the mobility changes with the external reorganization energy. In all cases the mobility monotonically decreases with external reorganization energy.

## III. SUMMARY AND CONCLUSION

In summary, we have calculated the charge carrier mobility (both electron and hole) in DPA-Tpa and Cpa-Cz dendrimers for both G2 and G3. We have shown that Cpa-cz has larger hole mobility than the electron while in Dpa-Tpa electron mobility is higher. The variation of mobility with respect to external electric field is also reported. We see that in CPA-cz dendrimer hole mobility decreases with the increase of electric field. At low electric field the G3 has higher mobility but after a critical field strength the mobility in G2 becomes higher. We have given physical explanation behind the observed electric field and generation dependence of mobility. Effect of the external reorganization energy on the mobility is also shown by calculating the mobility for various values of the external reorganization energy.

The dendrimers, we have studied have been already reported to increase the efficiency of DSSCs. However we hope that our investigation on the charge transport properties will induce more use of these dendrimers in optoelectronic device as well as DSSCs.

## ACKNOWLEDGEMENTS

We acknowledge financial support from DST, India.

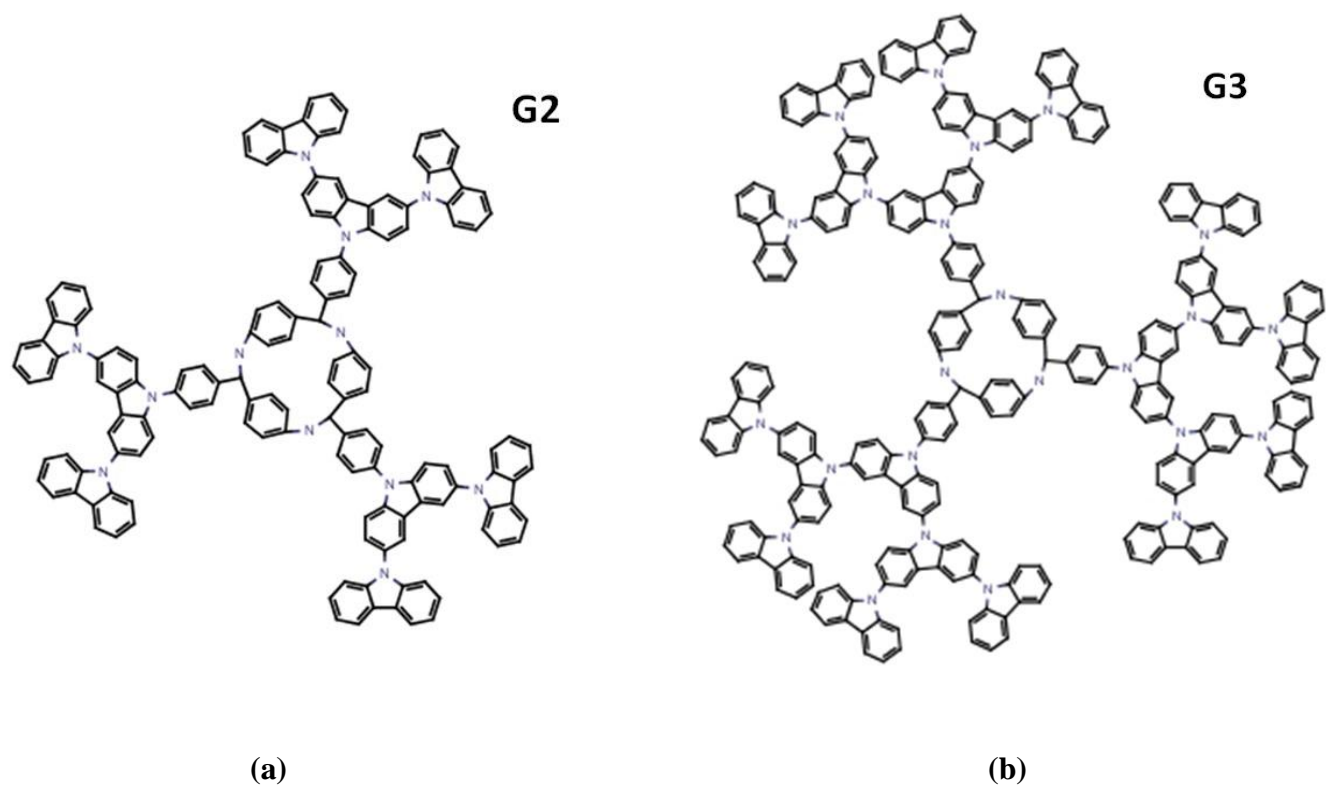

**(a)** **(b)**

Fig. 1. Molecular structure of the 2[nd] (a) and 3[rd] (b) generation Dendritic Carbazole with Cyclic Phenylazomethine as core (Cpa-cz). The structure consists of a central Cyclic Phenylazomethine core and carbazole units attached with the core.

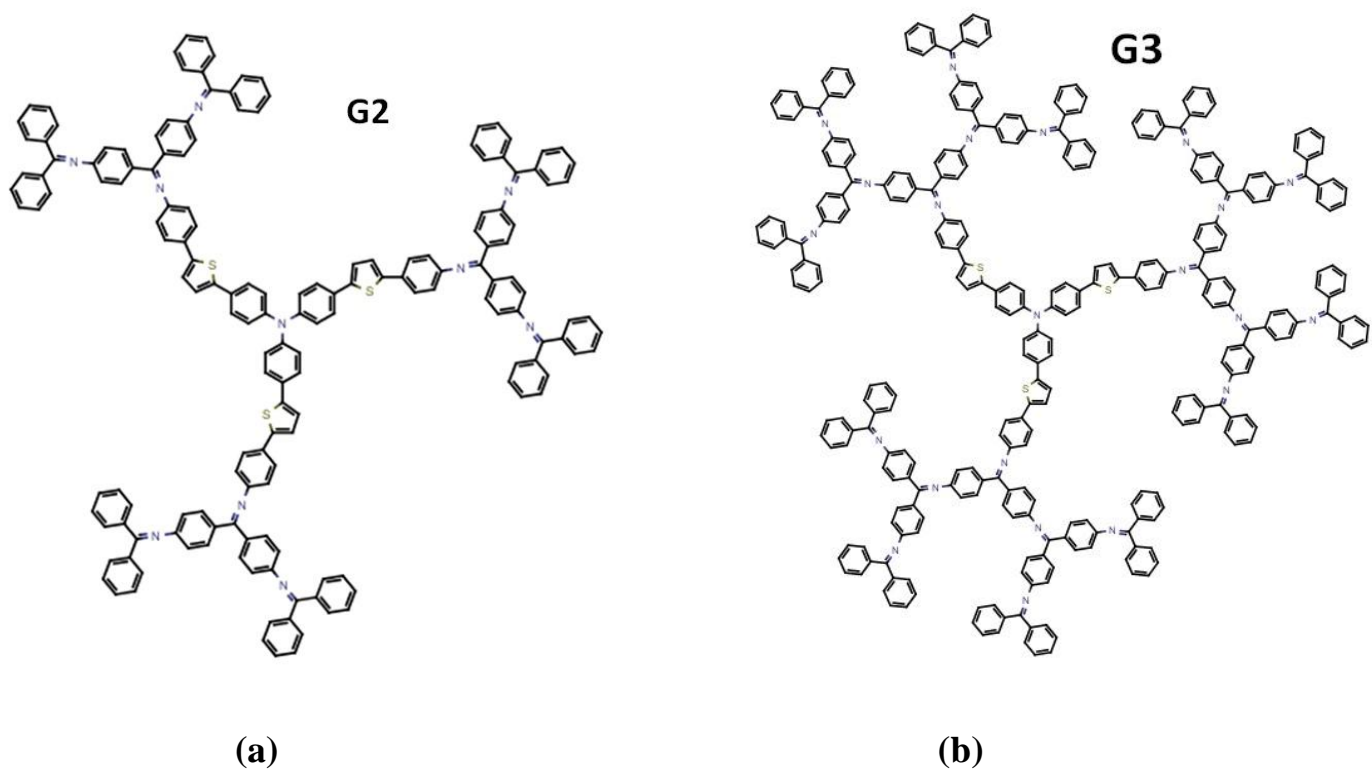

**(a)**                                                        **(b)**

Fig. 2. Molecular structure of the 2[nd] (a) and 3[rd] (b) generation of Dendritic phenyl azomethine with Triphenyl amine core (Dpa-Tpa). The structure consists of a central Triphenyl amine core and phenyl azomethine units attached with the core.

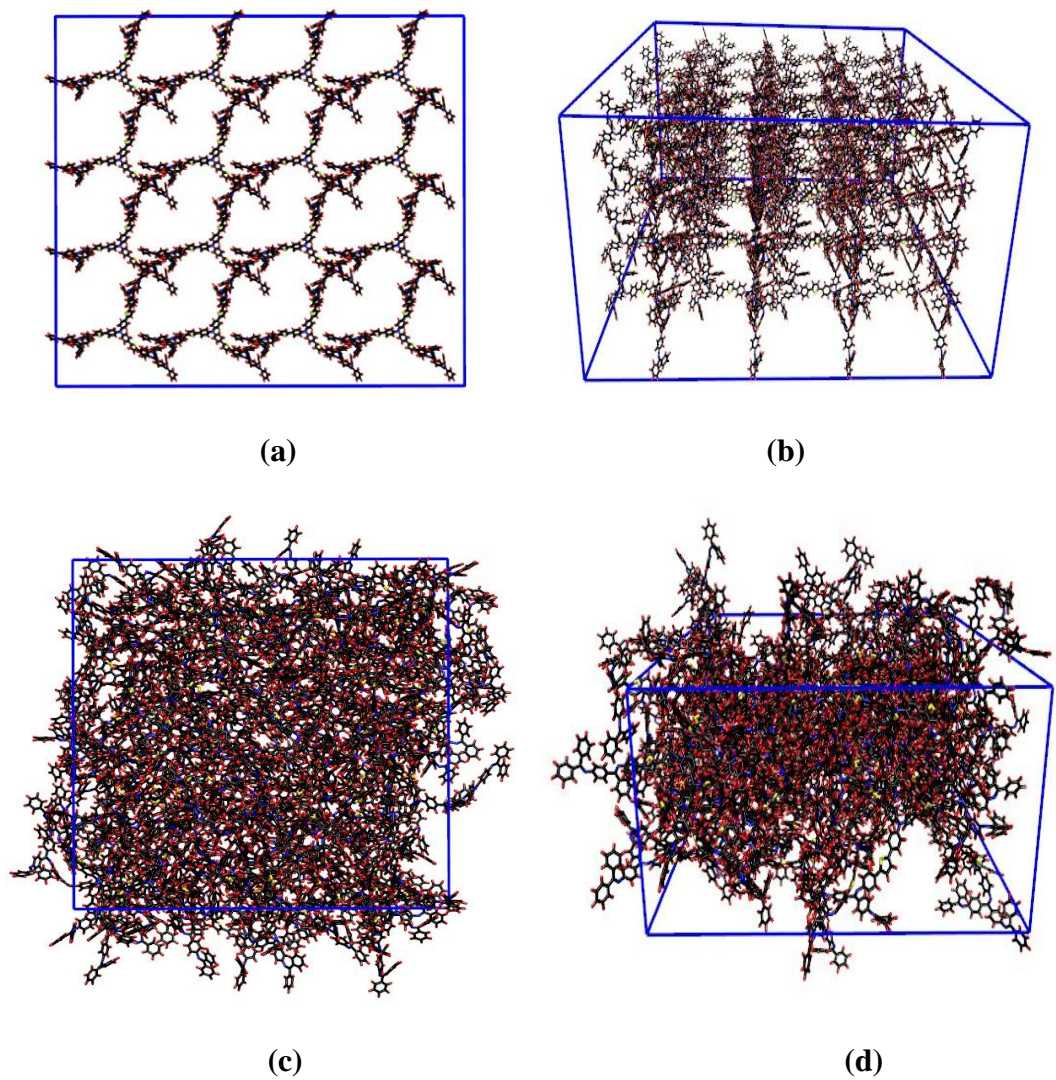

**(a)**　　　　　　　　　　　**(b)**

**(c)**　　　　　　　　　　　**(d)**

Fig. 3.  Snapshot of a system of 64 Dpa-Tpa (G2) molecules, prepared for simulation shown in top view (a) and side view (b) respectively. The molecules are arranged in 16 columns with 4 molecules in each column. An equilibrated representative simulation snapshot of the same system after 100 ns shown in top view (c) and side view (d) respectively.

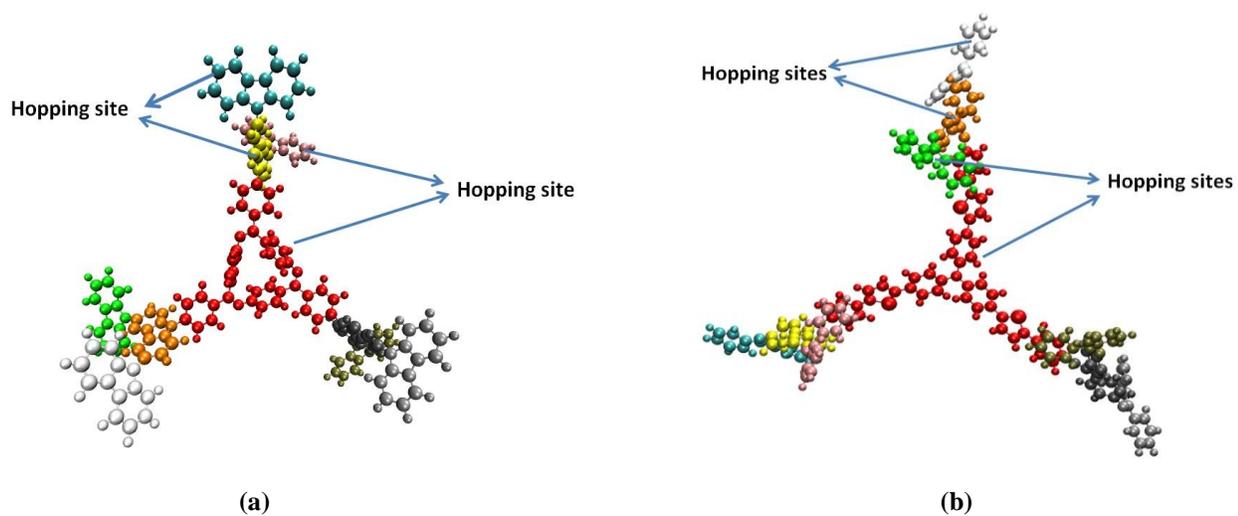

Fig. 4. Charge hopping sites of the 2<sup>nd</sup> generation (a) Cpa-cz dendrimer and (b) Dpa-Tpa dendrimer. Core and branches of the dendrimers are chosen as the charge hopping sites. All the atoms belonging to a hopping site are represented by same colour.

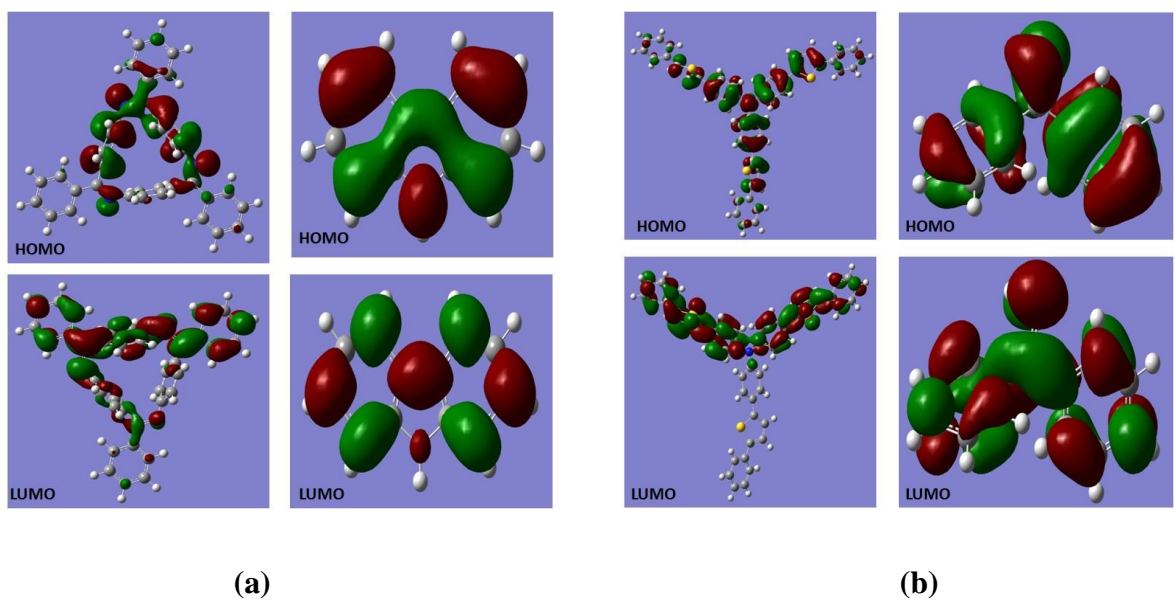

**(a)**                    **(b)**

Fig. 5. HOMO and LUMO orbitals of the core and branch of the CPA-Cz (a) and DPA-TPA dendrimers (b).

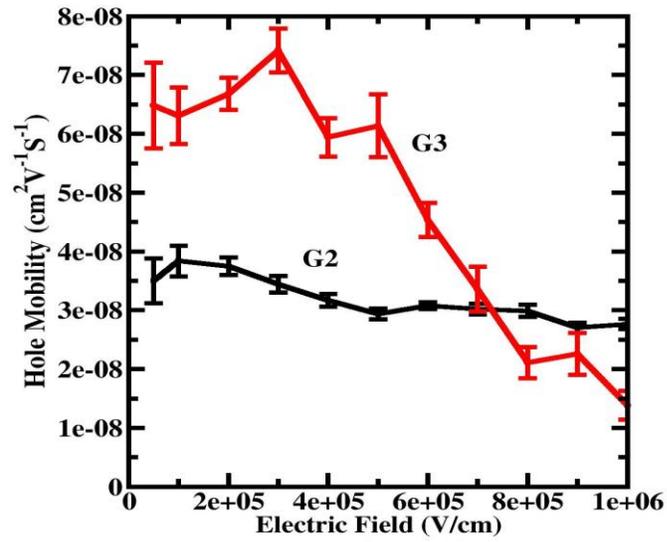

Fig. 6.  Hole mobility as a function of electric field strength in Cpa-Cz dendrimer melt system. At low field the mobility (hole) has higher value in G3 than G2 , but as electric field increases the mobility in  G3 falls down rapidly and after a certain electric field mobility in G2 takes over the mobility in G3.

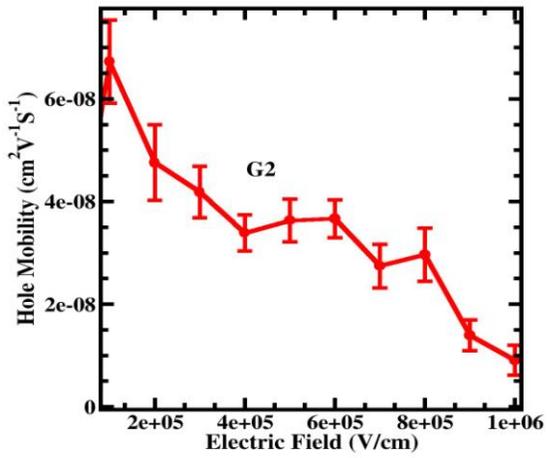
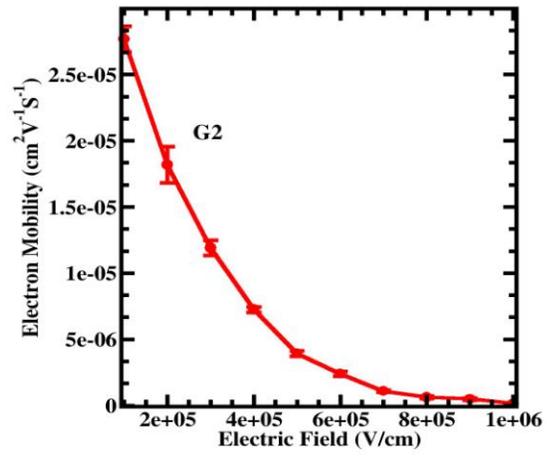

**(a)**　　　　　　　　　　　　　　　**(b)**

Figure 7:   Hole (a) and electron (b)  mobility as a function of electric field strength in 2$^{nd}$ generation Dpa-Tpa dendrimer melt system. Electron mobility is almost 3 order of magnitude higher compared to the hole.

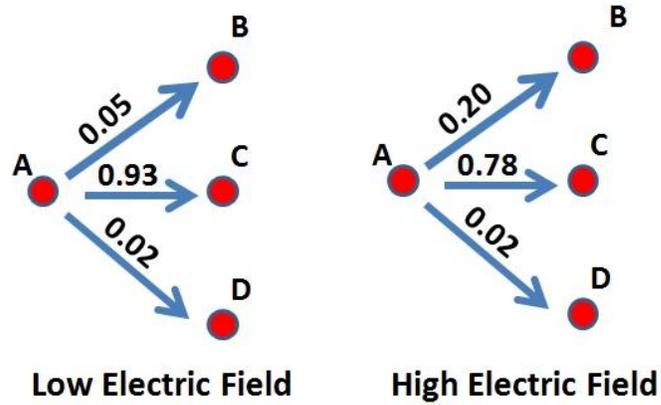

Figure 8: Schematic diagram representing few specific charge hopping sites and corresponding hopping probabilities for Cpa-cz (G3) dendrimer system. The red dots represent the hopping site. The blue arrows indicate the hopping events with the corresponding probability written along it. At low electric field the charge always hops from A to C while at high electric field there is finite probability for the charge to hop at site C as well as site B.

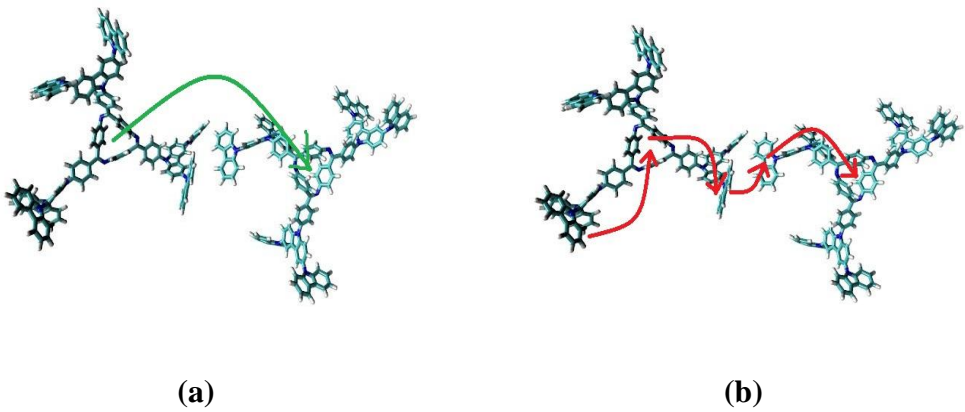

**(a)**                    **(b)**

Fig. 9. Two different charge transport pathways in dendrimer system. In first case (a), charge transport mainly happens through the cores of the dendrimers. In the other case (b), the charge hops from core to branch then  branch  to the branch of other dendrimer and in the end to the core of the latter.

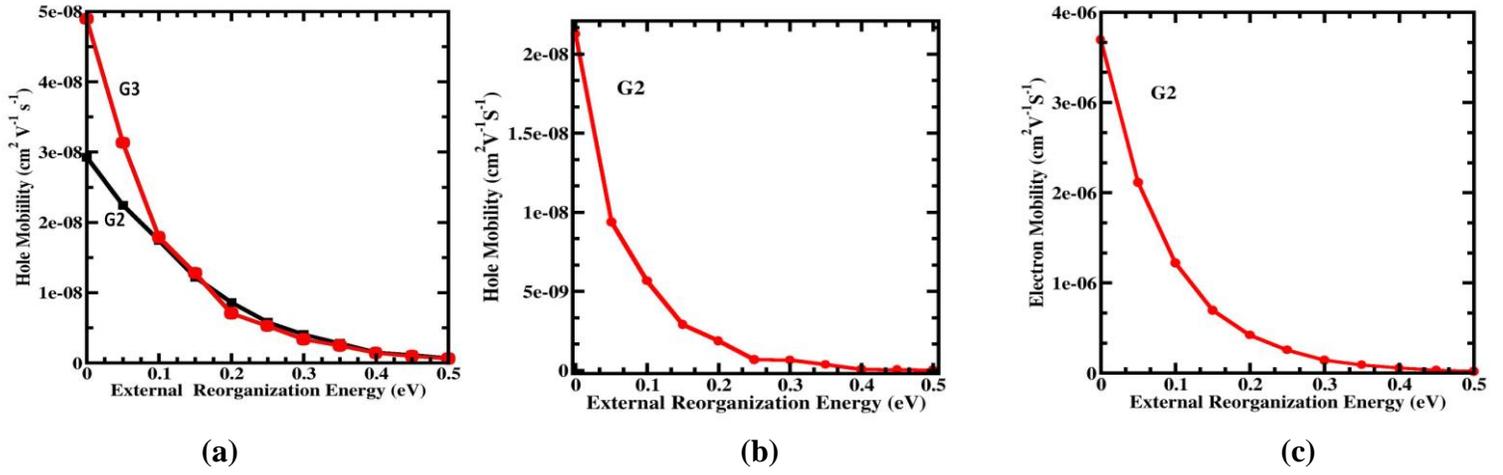

**(a)**                                      **(b)**                                      **(c)**

Fig. 10.  Hole mobility as a function of external reorganization energy in Cpa-Cz dendrimer melt system(a). Hole (b) and electron (c) mobility as a function of external reorganization energy for Dpa-Tpa dendrimer. Mobility decreases monotonically with the reorganization energy for all the cases.

# TOC GRAPHIC

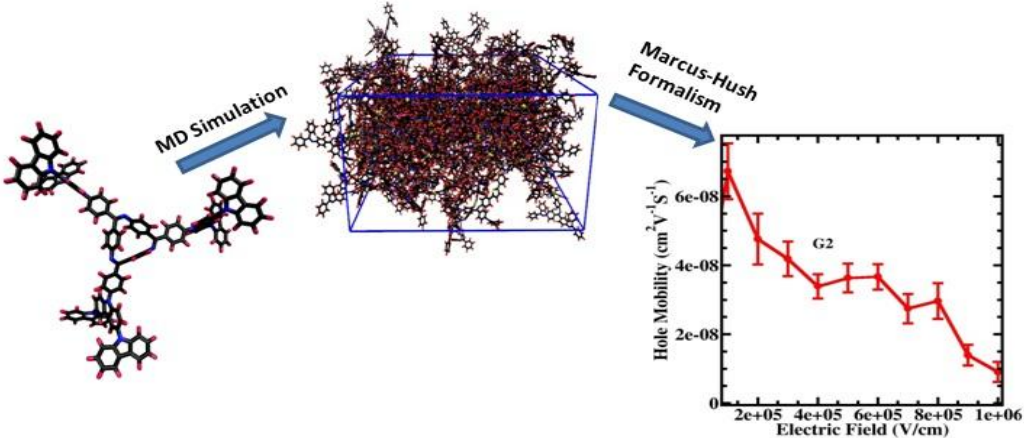